\documentclass[reprint,showpacs,amsmath,amssymb,floatfix,prb]{revtex4-1}

\usepackage{amsmath,amssymb,amsfonts}
\usepackage{graphicx,textcomp}
\usepackage{subfigure}

\newcommand{\figref}[1]{Fig.~\ref{#1}}
\renewcommand {\vec}    [1]    {\ensuremath{\mathbf{#1}}}

\begin{document}

\title{Thermal Anharmonic Effects in PbTe from First Principles}

\author{
A.H. Romero$^{1,2}$
}
\author{
E.K.U. Gross$^2$
}
\author{
M.J. Verstraete$^3$
}
\author{
Olle Hellman$^4$
}
\affiliation{
$^1$
Department of Physics, West Virginia University, 207 White Hall, 26506, WV, USA. \\
$^2$
Max-Planck-Institute f\"ur Mikrostrukturphysik, Weinberg 2, D-06120 Halle, Germany. \\
$^3$
Department of Physics, Universit\'e de Li\`ege, av du 6 ao\^ut, 17, B-4000 Li\`ege, Belgium\\
$^4$
Department of Physics, Chemistry and Biology (IFM), Link\"oping University, SE-581 83, Link\"oping, Sweden.
}

\begin{abstract}
We investigate the harmonic and anharmonic contributions to the phonon spectrum of lead telluride, and
perform a complete characterization of how the  anharmonic effects dominate
the phonons in PbTe as temperature increases. This effect is the strongest factor in the favorable thermoelectric properties of PbTe: an optical-acoustic phonon band crossing reduces the speed of sound and the intrinsic thermal conductivity. We present the detailed temperature dependence of the dispersion relation and compare our calculated neutron scattering cross section with recent experimental measurements. We analyze the thermal resistivity's variation with temperature and clarify misconceptions about existing experimental literature. This quantitative prediction opens the way to phonon phase space engineering, to tailor the lifetimes of crucial heat carrying phonons.
\end{abstract}

\maketitle

Heat conversion by using thermoelectric power generation has received a huge amount of interest in the last few years: transforming  a temperature gradient to a voltage difference promises to recover waste heat in thermal engines, transforming it into electrical energy. The thermoelectric efficiency of a material is captured by the figure of merit, $ZT=T S^2 \sigma/\kappa$, where $T$ is the temperature, $S$ is the Seebeck coefficient, and $\sigma$ and $\kappa$ are the electrical and thermal conductivities. The search for good thermoelectrics is centered on finding materials with a high figure of merit, which implies large electric and small thermal conductivities. Since the 1990s, a sequence of new materials which offer new paradigms in this field and a large number of energy harvesting applications have been proposed.~\cite{Zebarjadi2012, LaLonde2011a, Minnich2009, Dresselhaus2009}

Lead telluride is an industrial standard and a reference high performance thermoelectric, considered to be one of the materials to beat with alternative paradigms such as nanostructuring or nanoalloying. The properties of PbTe in its halite structure have been well documented and have been the basis of a large set of investigations, both in its pristine and alloyed configurations. The value of ZT for the crystalline phase of PbTe is close to 1.4.~\cite{LaLonde2011b} Recent investigations have reported values above 1.5, which can be obtained by $n$- or $p$-doping, as well as by considering band gap engineering~\cite{LaLonde2011a,LaLonde2011b}.
This value can be further increased by alloying or nanostructuring. For example,  Hsu {\it et al} recently reported a large figure of merit, close to 2.2 at 800K, for nanostructured PbTe.~\cite{Hsu2004} Alloying and nanostructuring minimize the lattice thermal conductivity, and one must try to keep the electrical conductivity unaffected. A different approach to increasing the figure of merit is to enhance the electrical power factor by engineering the electronic band structure.~\cite{Jaworski2013,Pei2012,Zhang2012} The use of PbTe as a starting material is the common factor in a long series of such improvements of ZT.

In order to understand the behavior of composite based-PbTe or more complex materials, it is important we can explain the thermoelectric properties of pristine PbTe, and, in particular, its low thermal conductivity between 450 and 800K, which is the basis for its high efficiency at ambient conditions. For the crystalline system, the low thermal conductivity has been correlated to the presence of large anharmonic effects at the $\Gamma$ point and zone boundary, and very low speed of sound~\cite{LaLonde2011b, Delaire2011, Singh2011, An2008}. These effects have been measured experimentally, and studied theoretically to a certain degree, by comparing to quasi-harmonic phonon spectra~\cite{Singh2011,Delaire2011} and by using perturbation theory for the anharmonic interatomic force constants.~\cite{Shiga2012,Tian2012} 

In the following, we spotlight the fundamental reason for which PbTe has a large thermal anharmonicity, how this anharmonicity goes beyond the second order response in the inter-atomic force constants, and how the optical-acoustic branch crossing is removed with increasing temperature. We stress the importance of full anharmonicities to be able to describe the temperature dependence of neutron scattering data. Additionally, we demonstrate the strong dependence of the thermal resistivity with respect to volume, which corresponds to the well known dependence with respect to applied pressure. For this quantity we also clarify differences with previously reported results, where good comparison was claimed  but with the wrong experimental data.
 

There have been numerous theoretical studies on the lattice dynamics of PbTe. Recent papers have focused on demonstrating the sensitivity of the optical phonons to changes in volume~\cite{An2008} and temperature.~\cite{Shiga2012,Tian2012} There is, however, still no conclusive theoretical model for the huge anharmonic effects in the lattice dynamics of PbTe. Experimentally, the phonon spectrum shows a strong dependence on temperature. One of the most important features reported in PbTe, and in particular the longitudinal/optical crossing which occurs around 1/3 along the path from $\Gamma$ to X.
The coupling between the acoustic and transverse optical (TO) modes has been identified to correlate strongly with the thermal conductivity,\cite{Shiga2012} and there is a very strong anharmonic coupling between the transverse mode, which is ferroelectric, and the longitudinal acoustic mode, which carries most of the heat. The crossing reduces the group velocity of the acoustic branch, and hence the thermal conductivity.
There is, however, still no theoretical explanation of the anomalous stiffening of the optical mode with increasing temperature.
This issue has been raised in Ref~\cite{Delaire2011} and a large difference was found between calculations in the quasiharmonic approximation (QHA) and experimental values. Delaire et al. concluded that the anharmonicity comes mostly from higher order terms in the inter-atomic force constants.
They report inelastic neutron scattering data, where the temperature effects are shown to affect strongly the zone boundary by hardening the TO modes, such that the crossing between LA and TO is lifted as a function of temperature. This is the central factor which limits the performance of PbTe at higher temperatures. 

We begin by characterizing the failure of calculations within the quasi harmonic approximation.
It is common practice to describe phonon systems using model Hamiltonians, built as Taylor expansions of the full ion-electron system as a function of the ionic displacements. 
%
%
From the second order, the frequencies $\omega_{\vec{q}s}$ are obtained as a function of mode $s$ and wave vector $\vec{q}$. Higher order force constants can, through perturbation theory,\cite{wallace1998thermodynamics} give the phonon self energy $\Delta_{\vec{q}s}+i\Gamma_{\vec{q}s}$,
which depends on the interatomic force constants to third and fourth order.
%
%
%

\begin{figure}[ht]
\begin{center}
\includegraphics[width=\linewidth]{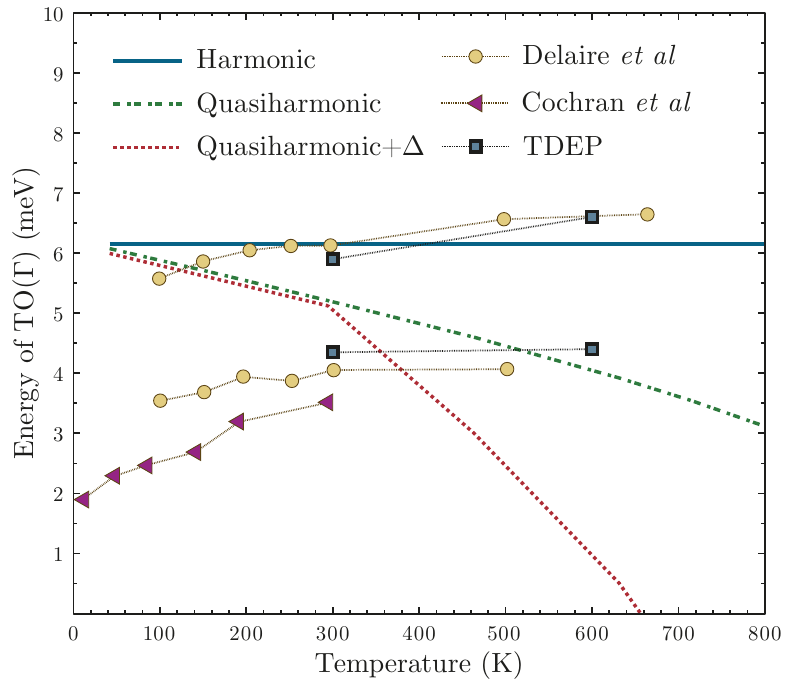}
\caption{\label{fig:tempevol_gamma}The energy of the TO($\Gamma$)-mode as a function of temperature. The harmonic line has no temperature dependence at all, the quasiharmonic has temperature dependence through the volume, the quasiharmonic+$\Delta$ is with the anharmonic shifts, which increases the disagreement. The experimental points are from  \cite{Delaire2011} (circles) and \cite{Cochran1966} (triangles).}
\end{center}
\end{figure}

Calculated Gr\"uneisen parameters in PbTe are all positive except for the TA modes. When temperature is increased, and the crystal expands, the optical modes should all soften in the quasiharmonic approximation, as reported in \cite{An2008}. In \figref{fig:tempevol_gamma} we show the energy of the TO modes at $\Gamma$ as a function of temperature. In the harmonic approximation, they are naturally constant. In the quasi-harmonic approximation, where $\Phi(T)=\Phi(V(T))$ and $V(T)$ is the volume as a function of temperature, the TO($\Gamma$) mode softens rapidly, as dictated by the $T=0$ Gr\"unesien parameters. If we add the anharmonic frequency shift, $\Delta_{\vec{q}s}(\omega_{\vec{q}s})$, the mode becomes unstable around 650K. This is all contrary to experimental trends: the TO($\Gamma$)-mode should stiffen with temperature, not soften. At the $\Gamma$ point several frequencies are seen in the neutron experiments, each corresponding to a peak in the scattering intensity. This is not reproduced at all in the QHA.

It is obvious by now that conventional quasiharmonic theory can not adequately describe PbTe at elevated temperature. Including the T=0 quasiharmonic self-energy shift $\Delta$ in the frequencies does not help, and even worsens the agreement, as seen in \figref{fig:tempevol_gamma}.

When a system is strongly anharmonic, second order model Hamiltonians, constructed at $T=0$,  
are not sufficient, and the Born-Oppenheimer energy surface has a nontrivial temperature dependence. To incorporate this in a lattice dynamical model we seek explicitly temperature-dependent force constants. We have performed extensive \emph{ab initio} molecular dynamics (AIMD) to sample the potential energy surface. Using the temperature-dependent effective potential technique (TDEP), we extract the temperature-dependent interatomic force constants~\cite{Hellman2011,Hellman2013,Hellman2013a}. 
Since AIMD contains phonon-phonon and electron-phonon coupling implicitly to all orders, the effective force constants will contain that information. They represent the best possible fit of the potential energy surface, for the current volume and temperature, using a finite order of force constants. 

\begin{figure*}[t!]
\begin{center}
\includegraphics[width=\linewidth]{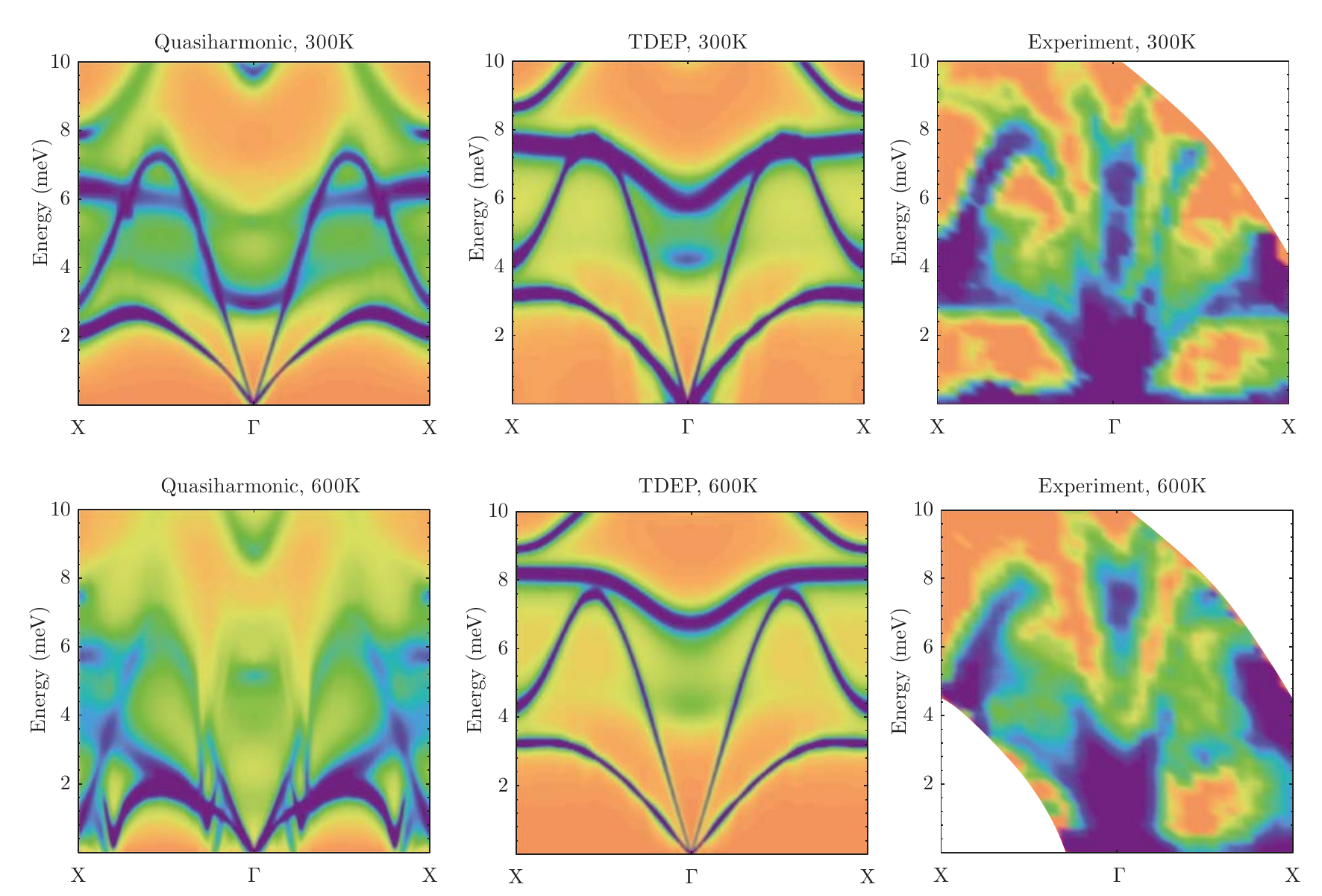}
\caption{\label{fig:pbte_sqe}PbTe Neutron scattering cross section. Notice the double peaks close to $\Gamma$ and the lifting of the optical branches above the acoustic captured within the TDEP method as compared to experiment. The quasiharmonic are included only for reference.}
\end{center}
\end{figure*}

We employ Born-Oppenheimer molecular dynamics in Density Functional Theory (DFT) with the projector-augmented wave (PAW) method as implemented in the VASP code.\cite{Kresse1999,Kresse1996,Kresse1993b,Kresse1996c} 
To converge the force constants fully, a $5 \times 5 \times 5$ supercell (250 atoms) is employed. For the BZ integration we use the $\Gamma$-point and run the simulations in the canonical ensemble, for a grid of temperatures and volumes. Temperature was controlled using a Nos\'e thermostat.\cite{Nose1984} Exchange-correlation effects were treated using the AM05 functional\cite{Armiento2005,Mattsson2009} and we use a plane wave cutoff of 250 eV. The simulations run for about 30ps after equilibration with a time step of 1 fs, ensuring proper ergodicity.
A subset of uncorrelated samples is then chosen, and for each of the samples the electronic structure and total energies are recalculated using a $3\times3\times3$ k-point grid and a cutoff of 300eV to obtain accurate forces.

Recent experimental results from \cite{Delaire2011} provide inelastic neutron scattering spectra for PbTe as a function of temperature. To compare with these experiments we convolve our data with the inelastic neutron scattering cross section:\cite{RACowley1968}

\begin{equation}
\!\!
\sigma_{\vec{q}s}(\Omega)
\propto
\frac{2\omega_{\vec{q}s}\Gamma_{\vec{q}s}(\Omega)}
{\left(\Omega^2-\omega^2_{\vec{q}s}-2\omega_{\vec{q}s}\Delta_{\vec{q}s}(\Omega)\right)^2
+4\omega^2_{\vec{q}s}\Gamma^2_{\vec{q}s}(\Omega)}
\end{equation}
\figref{fig:pbte_sqe} shows the theoretical and experimental inelastic neutron scattering cross section, the Debye-Waller factor and the experimental transfer functions have been disregarded. For the calculations of the self-energy, a $31 \times 31 \times 31$ $q$-point grid and a gaussian smearing of 0.1 meV have been used for the numerical evaluation of $\Gamma$ and $\Delta$.
The calculated and experimental spectra show remarkable agreement in absolute and relative shifts of the phonon frequencies. We accurately describe the double peak structure of the TO mode at $\Gamma$ at 300K. Traces of this double peak remain at 600K, which also agrees with the experimental picture. The energies of the two peaks as a function of temperature are shown with squares in \figref{fig:tempevol_gamma}, and agree well with the experimental values.

A crossing of the experimental acoustic and optical branches in the $\Gamma$-X direction is present at 300 K and disappears by 600K. This is also reproduced by our calculations. The quasi-harmonic results do not reproduce any of these features, and at 600K the self energy shifts have rendered the system completely unstable. Previous perturbation theoretical studies avoid this problem by disregarding either thermal expansion or $\Delta$.\cite{Shiga2012} From the quantitative agreement with experimental data, we conclude that we have built an accurate lattice dynamical model for PbTe at finite temperature with a good description of thermal properties and temperature dependencies. We now apply it to the calculation of the lattice thermal resistivity $\rho$. 
Experimental data for the thermal resistivity of PbTe has been measured in \cite{Devyatkova} and \cite{El-Sharkawy1983} at low and high temperature, respectively. 
In the literature \cite{Fedorov1969} is often referred to for the bulk lattice thermal conductivity of PbTe. This is not correct:  the paper does not contain any values of lattice thermal conductivity. The data attributed to Ref. \cite{Fedorov1969} is actually that contained in \cite{Devyatkova}. More importantly, close inspection of \cite{Devyatkova} shows that what is usually reported for the lattice thermal conductivity at high T is not measurement but a linear extrapolation from values obtained between 50 and 280K. The good agreement with experimental $\rho$ at high T in \cite{Shiga2012,Tian2012} is thus probably the result of the analytically linear behavior of $\rho$ in perturbation theory and the inappropriate linear extrapolation in \cite{Devyatkova}. The data of El Sharkawy \cite{El-Sharkawy1983} confirm the highly non-linear $\rho(T)$.

\begin{figure}
\begin{center}
\includegraphics[width=\linewidth]{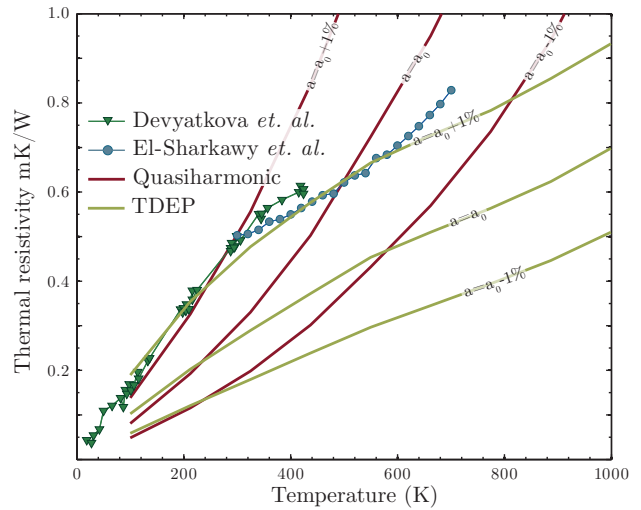}
\caption{\label{fig:pbte_thermalcond}(color online) Thermal resistivity of PbTe versus temperature. The red lines are QHA results with T=0 force constants and volumes corresponding to the experimental thermal expansion. The green lines are calculated using temperature-dependent force constants, with volumes following the experimental thermal expansion.}
\end{center}
\end{figure}

We calculate the resistivity from the phonon Boltzmann equation for fixed volume, beyond the relaxation time approximation and including isotope scattering as described in \cite{Broido2007}. We employ the same $31 \times 31 \times 31$ $q$-point grid and smearing of 0.1meV as previous perturbation calculations. The results are summarized in \figref{fig:pbte_thermalcond}, using the TDEP and quasiharmonic treatments, together with experimental results and considering different volumes for the theoretical calculations.

As temperature increases, the thermal resistivity increases slower than linearly. This is strongly correlated with the position of the TO mode with respect to the acoustic branches: the nested scattering between the LA and TO branches was already pointed out by Ref. \cite{Shiga2012}. When the TO mode is lifted up with increasing temperature, the available scattering volume shrinks dramatically, the acoustic phonons are allowed to carry heat unperturbed, and the resistivity is sub-linear with T.

At 100K both methods start in decent agreement with experiment. As T increases, however, the QHA results increase much faster, and the thermal expansion effect pushes the results away from the experimental curves. TDEP values are more reasonable and are smaller than experiment: additional possible scattering mechanisms (defects, grain boundaries) must \emph{add} to the resistivity. In particular the sub-linear behavior observed up to 600 K is reproduced correctly. At higher temperatures the expected linear behavior is restored, due to the high T linear limit of the Bose Einstein distribution for each given frequency.

Additionally, perturbation theory becomes questionable in the case of PbTe: for a few modes, the linewidth and frequency can be comparable and $\Gamma_{\vec{q}s} \ll \omega_{\vec{q}s}$ definitely does not hold. What our results suggest is that the TDEP renormalized quantities \emph{do} obey the relation $\Gamma_{\vec{q}s}(T) \ll \omega_{\vec{q}s}(T)$. Finally, the large Gr\"uneisen parameters imply a strong volume dependence. In \figref{fig:pbte_thermalcond} we show the thermal resistivity for lattice parameters compressed and expanded by 1\%, an expansion value that can be estimated from the experimental thermal expansion. This leads to a factor 2 difference in resistivity. The intrinsic exchange-correlation error in current DFT functionals is certainly of this order of magnitude for the volume, and more accurate calculations will necessitate further improvements in the fundamentals of DFT.

Our results show clearly that a careful examination of the phonon response is crucial, when dealing with systems as strongly anharmonic as PbTe. The TDEP method contains the anharmonic terms necessary to describe phonon frequencies at high T. Quantitative prediction of the exceptionally low thermal resistivity is possible but still  very challenging in this delicate system.
Many successful but heuristic attempts have been made to further reduce the low thermal conductivity of PbTe (by doping, alloying and nanostructuring). We show here that a direct rational approach, based on first principles calculations, can now be taken. Exploiting the volume (and hence stress) dependency of the phonon frequencies, the phonon resistivity will be maximized by preserving the optical-acoustic crossings. Many other high performance thermoelectrics present low-lying optical modes, and could be manipulated in similar ways: using epitaxy to start from a lower lattice constant at ambient, or uniaxial stress to create a preferred direction for the heat flow. 
It is common to consider phonon band engineering, invoked in many nanostructuring or nanoalloying strategies, to reduce the speed of sound and break the group velocity of acoustic modes, e.g. using disorder or rattlers. Much less is quantified about the reduction of phonon \emph{lifetimes} by manipulating the available phase space for decay, as we have demonstrated happens in PbTe, confirming the work of Shiga \emph{et al.} \cite{Shiga2012}.
The presence of anharmonicity is intimately linked to the mass of the atoms and overall softness of the phonon modes, which are important features of most successful thermoelectrics.

Summarizing, we describe the relevance of high orders of the inter atomic force constants in the proper description of the thermoelectric properties of PbTe. This result should guide future work on alloy-based thermoelectrics as well as other high temperature applications. These features will also be present in similar compounds, where strong anharmonicities have been reported, and can not be accounted for by normal perturbation theories.  



\begin{acknowledgments}
AHR acknowledges the support of the Marie Curie Actions from the European Union in the international incoming fellowships (grant PIIFR-GA-2011-911070). MJV acknowledges an ARC grant (TheMoTherm \# 10/15-03) from the Communaut\'e Fran\c{c}aise de Belgique, and computer time from CECI, SEGI, and PRACE-2IP (EU FP7 Grant No. RI-283493) on Huygens and Hector.
\end{acknowledgments}

\bibliographystyle{pnas}
\bibliography{library}

\end{document}